\documentclass[journal]{IEEEtran}
\usepackage{amsmath}

\usepackage{enumitem}
\usepackage{graphicx}
\usepackage{caption}
\usepackage{stfloats}
\usepackage{multirow}
\usepackage{array}
\usepackage{hyperref}
\usepackage{subfig}
\usepackage{cite}

\ifCLASSINFOpdf
\else
\fi

\begin{document}
\title{Feature Aggregation in Joint Sound Classifcation\\ and Localization Neural Networks}
\author{Brendan~Healy,
        Patrick~McNamee,
        and~Zahra~Nili~Ahmadabadi~(\IEEEmembership{IEEE~Member})

\thanks{\textit{Corresponding author: Zahra Nili Ahmadabadi}}
\thanks{Brendan Healy is with the Mechanical Engineering Department at San Diego State University, San Diego, CA 92182 USA (e-mail: brendanhealy408@gmail.com).}
\thanks{Patrick McNamee is with the Mechanical Engineering Department at San Diego State University, San Diego, CA 92182 USA (e-mail: pmcnamee5123@sdsu.edu).}
\thanks{Nili Ahmadabadi is with the Mechanical Engineering Department at San Diego State University, San Diego, CA 92182 USA (e-mail: zniliahmadabadi@sdsu.edu).}}


\maketitle
\begin{abstract}
This study addresses the application of deep learning techniques in joint sound signal classification and localization networks. Current state-of-the-art sound source localization (SSL) deep learning networks lack feature aggregation within their architecture. Feature aggregation enhances model performance by enabling the consolidation of information from different feature scales, thereby improving feature robustness and invariance. This is particularly important in SSL networks, which must differentiate direct and indirect acoustic signals. To address this gap, we adapt feature aggregation techniques from computer vision neural networks to signal detection neural networks. Additionally, we propose the Scale Encoding Network (SEN) for feature aggregation to encode features from various scales, compressing the network for more computationally efficient aggregation.  To evaluate the efficacy of feature aggregation in SSL networks, we integrated the following computer vision feature aggregation sub-architectures into a SSL control architecture: Path Aggregation Network (PANet); Weighted Bi-directional Feature Pyramid Network (BiFPN); and SEN. These sub-architectures were evaluated using two metrics for signal classification and two metrics for direction-of-arrival regression. The results suggest that models incorporating feature aggregations outperformed the control model, the Sound Event Localization and Detection network (SELDnet), in both sound signal classification and localization. Among the feature aggregators, PANet exhibited superior performance compared to other methods, which were otherwise comparable. The results provide evidence that feature aggregation techniques enhance the performance of sound detection neural networks, particularly in direction-of-arrival regression.
\end{abstract}

\begin{IEEEkeywords}
Joint sound signal classification and localization, Multi-task deep learning, Feature aggregation 
\vspace{\baselineskip}
\end{IEEEkeywords}

\noindent\textbf{Availability of data, material, or code:}

\url{https://gitlab.com/dsim-lab/paper-codes/feature-aggregation-for-neural-networks}

\IEEEpeerreviewmaketitle

\section{Introduction}
\IEEEPARstart
{S}{ound} source localization (SSL) represents an imperative domain within the broader field of audio signal processing, holding significant implications for topics such as robotics, hearing aids, and speech recognition systems \cite{chakrabarty2019broadband}. SSL techniques aim to ascertain the location or direction-of-arrival (DOA) of a sound source, which provides critical data for sound source separation \cite{chazan2019multi}, speech augmentation \cite{xenaki2018sound}, robot-human interaction \cite{li2016reverberant}, noise control \cite{chiariotti2019acoustic}, and auditory scene analysis \cite{grumiaux2022survey}.  

A key gap within existing SSL neural networks is the lack of feature aggregation within their architecture. Feature aggregation can boost a model performance by consolidating information from various scales and contexts, thereby enhancing feature robustness and scale invariance. It is particularly vital for SSL networks, which must distinguish between direct signals and reflections \cite{adavanne2018sound}. This paper aims to address this gap by adapting feature aggregation techniques from computer vision neural networks and applying them to signal detection neural networks. Moreover, we propose the development of a novel architecture, the Scale Encoding Network (SEN), which serves as a compact feature aggregator in the context of SSL.

\subsection{Related Work}
Early endeavors in machine learning for SSL were focused on conventional machine learning models, namely the Multilayer Perceptron (MLP) and Support Vector Machines (SVM) \cite{rasheed2013sound, jiang2014novel}. The aforementioned models encountered difficulties, particularly in effectively managing large datasets and addressing the complexities associated with temporal relationships in the input features. In light of these difficulties, there has been a notable shift towards Convolutional Neural Networks (CNNs), which have demonstrated the ability to capture spatial features in data \cite{grumiaux2022survey, adavanne2018deep}. As deep learning techniques have advanced, the development of Recurrent Neural Networks (RNNs) gave rise to Convolutional Recurrent Neural Networks (CRNNs). This combination successfully utilized both spatial and temporal dimensions of the data, leading to improved DOA estimation \cite{grumiaux2022survey, xie2019doa}.

Residual CNNs (Res-CNNs) soon emerged, incorporating shortcut connections that provide a link between the input and output layers. Res-CNNs proved superior in performance compared to both conventional CNNs and CRNNs \cite{zhang2020residual, he2016deep}. The introduction of novel architectures such as Res-CRNNs and deep generative models enhanced the DOA estimation \cite{goodfellow2014generative, ranjan2019sound}. Additionally, attention mechanisms have continued to enhance the capabilities of neural networks by allowing them to selectively concentrate on pertinent features, improving the accuracy of the estimation process \cite{kim2018voice}.

\subsection{Contributions}
This paper makes the following contributions:

\begin{enumerate}
  \item We introduce Feature Aggregation techniques from image based Object Detection Neural Networks into SSL and Sound Detection Res-CRNN networks. 
  
  \item We propose and test a new Feature Aggregator method, the SEN.
  
  \item We provide a publicly available Feature Aggregator Library for TensorFlow’s Functional API. This library contains pre-made aggregators and allows for the efficient creation of new aggregators.
\end{enumerate}

\subsection{Overview}
The remainder of this paper is organized as follows. Section \ref{sec:theory} explains challenges with feature scaling in SSL networks, the role of feature aggregation in resolving them, and its practical application in real-world models. Section \ref{sec:method} elaborates on the training and testing procedures used to gather data for evaluating the merit of our theory. Section \ref{sec:eval} describes the dataset, preprocessing, evaluation metrics, and baseline methods used to gauge our results. Section \ref{sec:res} investigates the testing outcomes and their contextual meaning. Section \ref{sec:conc} provides a summary of our findings, their implications, and directions for future research.

\section{Theory} \label{sec:theory}
This section is divided into four parts. Section \ref{sec:theory}.A discusses the importance of scale invariance within neural networks. Section \ref{sec:theory}.B explains how feature aggregation addresses scale invariance, how feature aggregation is performed, and various feature aggregation designs. Section \ref{sec:theory}.C elaborates on our custom aggregation approach. Section \ref{sec:theory}.D gives an overview of standard object detection designs and how feature aggregation is employed within a full architecture.

\subsection{Scaling}
Feature scaling is a powerful tool in both object detection and SSL neural networks. The ability to learn notable patterns in data and then identify these patterns at different scales reduces the amount of training data required and chances of overfitting to a particular size or amplitude input. The scale of extracted features in CNNs depends on the tensor dimensions and convolutional hyperparameters used in each layer of the network \cite{zeiler2014visualizing, yosinski2014how}. As the input tensor is passed through the network, features are extracted at different scales. Downsampling, such as max-pooling or strided convolutions, reduces the size and spatial resolution of the feature map. As a result, finer resolution features that are present in earlier network layers may be neglected in subsequent coarser resolution layers. This phenomenon is known as the "semantic gap," where the features in different layers of the network represent different levels of abstraction, and finer features may be ignored as the network learns more complex representations \cite{zeiler2014visualizing, yosinski2014how}.

This semantic gap is of extreme importance in perception models because features of smaller scale can be overlooked in deeper convolutions. For example, in computer vision, features from distance or small objects may be lost as the input image is downsampled throughout multiple convolutions. This means the model loses the ability to identify a class at various sizes and distances and therefore is quite limited in its uses. In real- world applications, most classes vary in size and will not be at the same exact location in each image. 

This same principle applies to SSL and may be even more important. This is because SSL algorithms must differentiate between direct and indirect signals (such as reflections, reverberations, and diffractions). These indirect signals generally have a similar (or identical) wave pattern as their source’s direct signals, but at a reduced amplitude and with a phase shift. From a neural network's perspective, features of quieter or indirect signals are the equivalent to another source that is further away; the distinguishing patterns are the same but of different amplitude and phase. To differentiate between direct and indirect signals, a SSL model should: 1) identify all signals from the same source; and 2) isolate the direct signal based on its scale relative to the indirect signals. For both these steps, the model must understand feature scaling. 

To address the semantic gap, specific architectures have been proposed, such as U-Net and Feature Aggregators. Various studies have conducted performance comparisons between Feature Aggregation and U-Net architectures, demonstrating that feature aggregation networks achieve high performance in various image segmentation tasks \cite{liu2018auto}. Although, tasks such as brain tumor segmentation, which have more emphasis on fine-grained details, benefit  from using a U-Net which has been trained on a sufficiently large dataset.\cite{isensee2021nnunet, ronneberger2015unet}. Additionally, feature aggregation networks exhibited enhanced computational efficiency, reduced memory footprint, and the ability to achieve high accuracy with smaller training datasets. The latter was demonstrated in a study that examined the effectiveness of feature aggregation networks in the context of image segmentation \cite{han2020feature}. The demand for less data is particularly crucial in the context of sound detection models as each class of signal must be sampled at many angles or locations; with variables like room size/shape, wall material, and objects in the room affecting signal reflections and reverberations.

Therefore, the choice between feature aggregation and encoder architectures ultimately depends on the desired tradeoff between efficiency and accuracy. Some tasks may require more emphasis on fine-grained details and thus benefit from the use of U-Net, while others may prioritize computational efficiency and simplicity, making feature aggregation a more suitable option \cite{ronneberger2015unet}. This study focuses on feature aggregation due to its practicality in real world applications. Large computational cost, memory footprint, and training dataset requirements make sound detection U-Nets impractical for sound detection.

\subsection{Feature Aggregation}
The purpose of feature aggregation is to combine features from various convolutions throughout the network to improve scaling, overfitting, and exploding or vanishing gradients \cite{tan2019efficientdet}.

\begin{figure}
    \centering
    \includegraphics[width=0.3\textwidth]{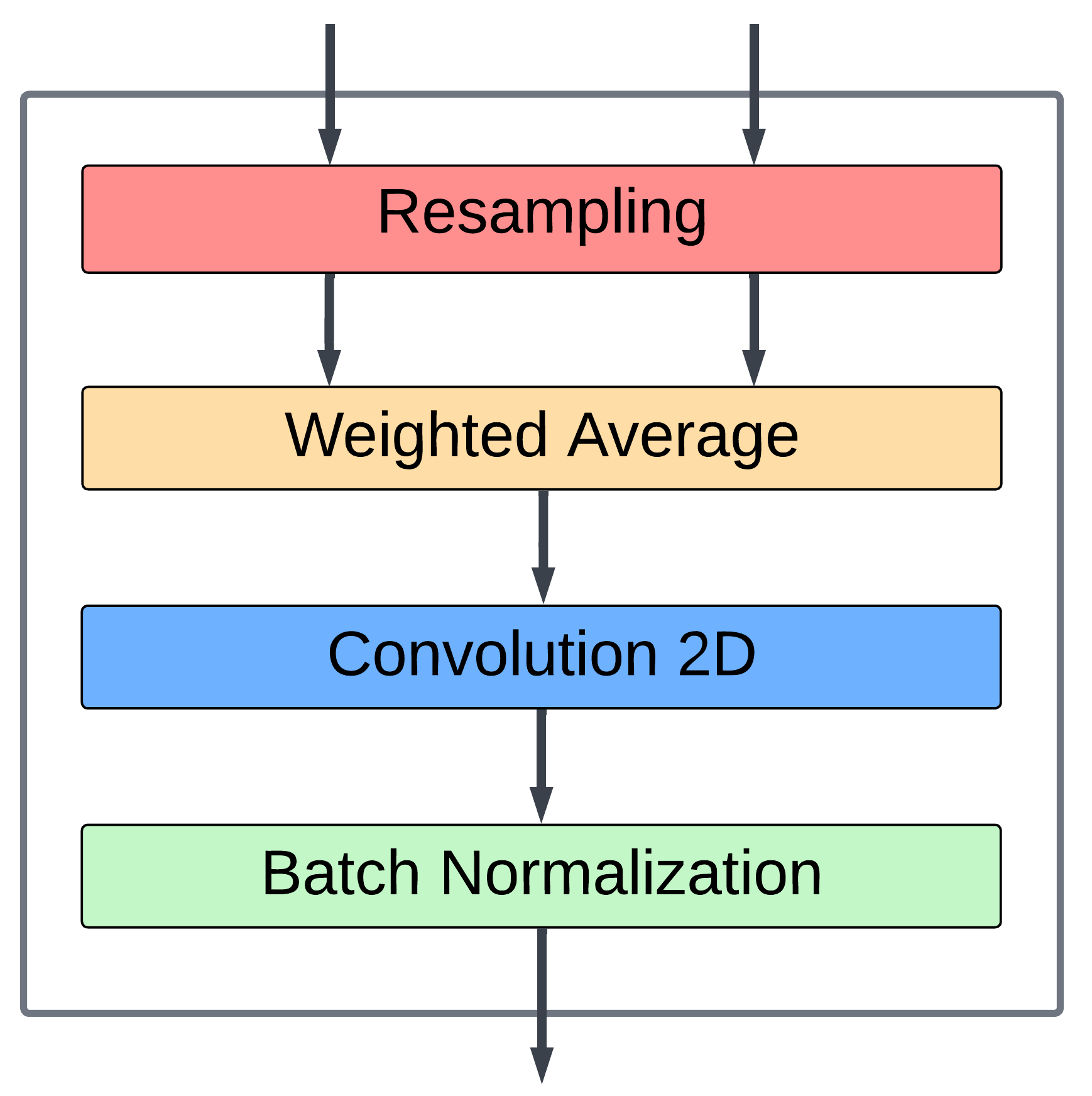}
    \caption{Aggregation node \cite{tan2019efficientdet} diagram illustrating sequential procedures performed within each node.}
    \label{fig:Node Diagram}
\end{figure}

Feature aggregation has three sequential steps: resampling inputs to match shapes, aggregate inputs (weighted averaging or concatenation), and convolution of the aggregated tensor \cite{tan2019efficientdet}. These processes are completed inside a TensorFlow sub-model called a “Node”, as is illustrated in Fig. \ref{fig:Node Diagram}. As is discussed later in this section, multiple nodes are connected residually within an aggregator, allowing this process to be repeatedly executed throughout an elaborate structure.

Resampling consists of two processes, downsampling and upsampling \cite{tan2019efficientdet}. Downsampling reduces the resolution of feature maps, resulting in a smaller spatial dimension but a larger number of channels. It is typically done through operations such as max pooling or strided convolutions. In contrast, upsampling increases the resolution of feature maps by interpolating the values and can be achieved through techniques such as deconvolutions, transposed convolutions, or interpolation \cite{tan2019efficientdet, redmon2016yolo, liu2016ssd}. One commonly used method for upsampling tensors in computer vision and image processing is bilinear interpolation. It is used to compute the value of a new pixel (i,j) by taking the weighted average of the values of its four nearest pixels in the original image \cite{rogers1990mathematical}. 






Bilinear interpolation is preferred over other interpolation methods like nearest-neighbor interpolation because it produces smoother and more natural-looking results \cite{long2015fully}. Nearest-neighbor interpolation simply selects the nearest pixel value without considering its neighboring pixels, which can result in jagged edges and other artifacts \cite{redmon2016yolo}. This spatial preservation of features is vital for the resampling of both images and spectrograms \cite{adavanne2018sound, adavanne2018deep}. Additionally, bilinear interpolation can be easily extended to higher dimensions, such as 3D volumes or tensors \cite{han2020feature}. It is also relatively easy to implement and understand, making it a popular choice for spatial features \cite{liu2018auto}. This study utilizes bilinear interpolation over other upsampling methods due to the reasons previously listed.

Industrial computer vision object detection models that demonstrate resampling are YOLO (You Only Look Once) \cite{redmon2016yolo} and SSD (Single Shot MultiBox Detector) \cite{liu2016ssd}. In YOLO, downsampling is performed using strided convolutions, while upsampling is achieved using transposed convolution. In SSD, downsampling is performed using max pooling, and upsampling is done using deconvolution layers. Comparatively, this study performs downsampling using strided convolutions and upsampling through bilinear interpolation. Our choice for resampling methods is based on \cite{tan2019efficientdet}; a study that examines the efficiency of aggregation methods in computer vision and uses bilinear interpolation during its more efficient aggregation. The difference in resampling approaches demonstrates that there is not one universal method for the resampling process. The choice of algorithm for aggregation steps can vary depending on desired computational cost and desired performance.

Once resampling is complete, the tensors are aggregated using weighted averaging. This is more efficient than the conventional method of concatenation as it results in a smaller tensor \cite{brandstein2001microphone}. Weighted averaging allows the model to directly contrast features derived at distinct scales and processing depths, diversifying the scale and refinement level of features used for final predictions \cite{lin2017feature}. The weights are trainable variables, allowing the model to optimize the aggregation.

\begin{figure}
    \centering
    \includegraphics[width=0.45\textwidth]{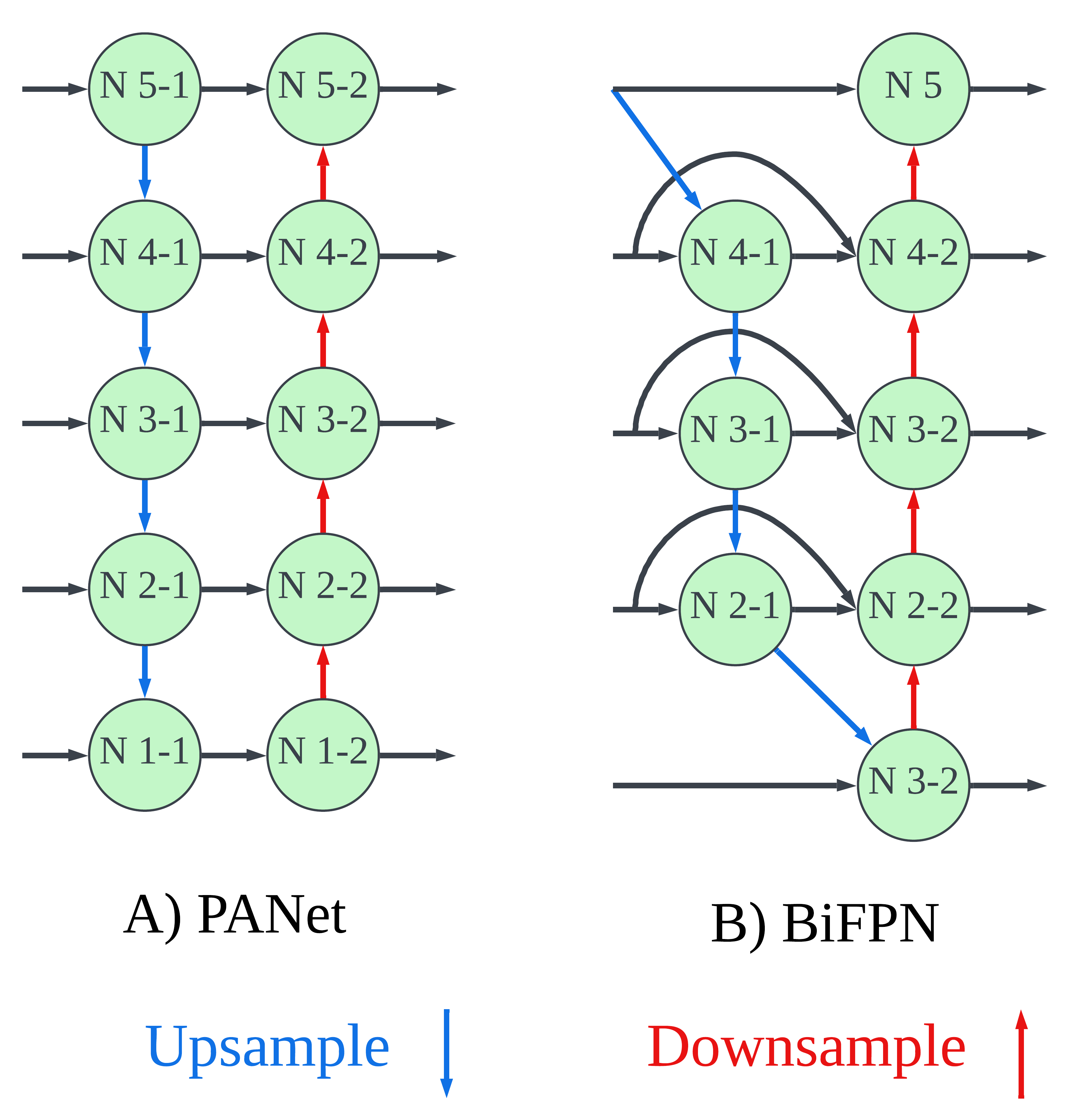}
    \caption{Example diagrams of PANet \cite{liu2018path} and BiFPN \cite{tan2019efficientdet} feature aggregators with five scales.}
    \label{fig:PANet + BiFPN Diagram}
\end{figure}

As seen in Fig. \ref{fig:PANet + BiFPN Diagram}, feature aggregators connect nodes via residual connections, creating a complex residual network. A benefit of these residual connections is the minimization of vanishing/exploding gradients throughout the network \cite{lin2017feature}. Residual connections enable the gradient to flow through the network more easily, stabilizing the training process. In Path Aggregation Network (PANet), residual connections propagate high-level semantic features from the top-down pathway to the bottom-up pathway, enabling the network to generate highly detailed object proposals at multiple scales \cite{lin2017feature, liu2018path}.

Feature Aggregators are classified based on their number of nodes and connection order, which varies with the feature extraction depth and structure. However, aggregation of additional feature scales results in computational overhead, necessitating a tradeoff between higher aggregation and prediction speed \cite{tan2019efficientdet, lin2017feature}. This is why recent object detection models transitioned from Feature Pyramid Network (FPN) to PANet for enhanced performance \cite{liu2018auto}, yet subsequent developments have focused on more compact alternatives such as Neural Architecture Search Feature Pyramid Network (NAS-FPN) \cite{ghiasi2019nasfpn} and Weighted Bi-Directional Feature Pyramid Network (BiFPN) \cite{tan2019efficientdet}.

\subsection{Scale Encoder Network}
The novel aggregator introduced in this study is Scale Encoder Network (SEN). The goal of SEN is to reduce aggregation’s computational complexity from the node aggregation process while still enabling the neural network to weigh scales in a residual manner. Its premise is based on the idea that encoding multiple scales into one scale results in less nodes but still addresses the semantic gap.

\begin{figure}
    \centering
    \includegraphics[width=0.35\textwidth]{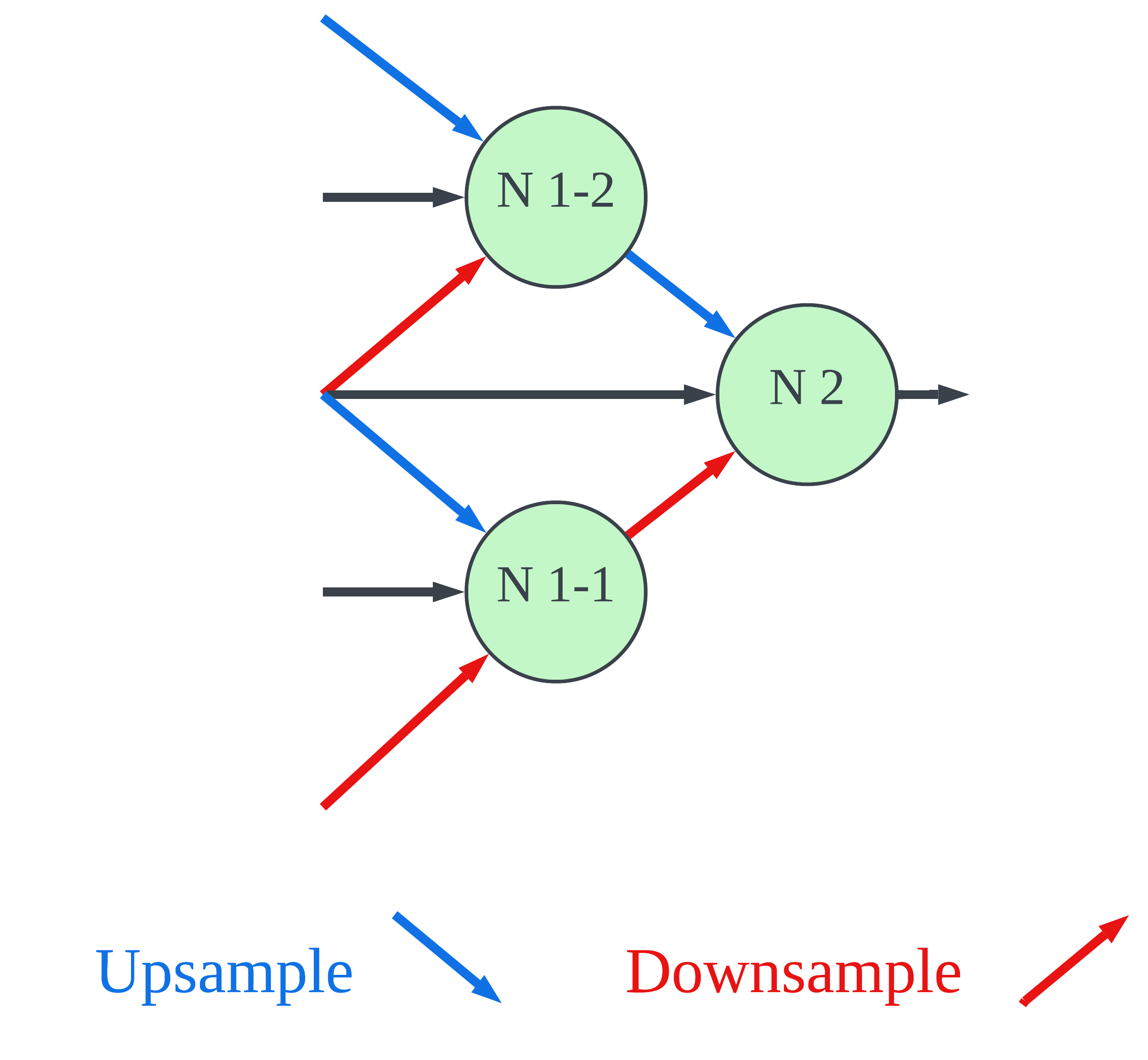}
    \caption{Example diagram of SEN feature aggregator encoding five scales down to one scale.}
    \label{fig:SEN}
\end{figure}

Aggregators such as FPN, PANet, and BiFPN essentially update a constant number of scales throughout their process. If the backbone of the network has N nodes (or scales), the final layer of these aggregators also has N nodes and outputs. SEN, on the other hand, compresses multiple scales into one, as seen in Fig. \ref{fig:SEN}. In SEN, consecutive aggregation layers reduce N until it is equal to the desired number of outputs. In Fig. \ref{fig:SEN}, five initial scales are compressed into two nodes, then one node during aggregation.

In \cite{tan2019efficientdet}, seven backbone outputs feed into aggregators. In this case, PANet adds 14 resamplings, weighted averages, and convolutions to a network of only seven convolutions. A SEN with a compression width of two would have a first layer of three nodes and a second layer of one node. In models like DarkNet \cite{redmon2018yolov3}, there can be over 50 convolutional blocks in the backbone, and a SEN aggregator will have a huge impact on the computational cost of feature aggregation.

The number of scales between nodes in a SEN layer is referred to as the compression width; there are several factors to consider when choosing it. The first factor is the volume and scope of resampling. While downsampling results in data loss, upsampling necessitates data approximation. In many feature aggregators (PANet, FPN, BiFPN, NAS-FPN), scales are never resampled more than one scale at a time \cite{tan2019efficientdet, ghiasi2019nasfpn}. Presumably, this is to minimize data loss and approximation throughout the network. In an attempt to not resample one scale too much, we chose to use the middle backbone scale for SEN outputs. Connection overlaps should also be considered; a compression width of one can result in many overlapping connections, which could lead to repetitive calculations and a preference for particular scales. To compare the effects of various compression width sizes, this study evaluates two different SEN designs; which are explained further in Section III.

\subsection{Object Detection Architecture}
This section discusses Multi-Task Learning (MTL) as well as the various stages of object detection in neural networks. These aspects of neural network architecture are not exclusive to object detection networks but are pivotal for object detection tasks. MTL refers to the process of training neural networks to make predictions for multiple tasks, such as object localization and classification, simultaneously \cite{caruana1997multitask}. This enables models to train variables while also cross-referencing losses from each task. By collectively downplaying noise and anomalous patterns that one task may have overemphasized, each task contributes evidence for the applicability of features. This prevents overfitting by focusing on crucial traits. Eavesdropping occurs when information obtained from simple tasks is used to finish a complex task \cite{caruana1997multitask}. For instance, filters used in image segmentation to identify an object's class provide information about the object's shape, which can be used to calculate the object's coordinates.

Usually, object detection models consist of the three steps of feature extraction, feature aggregation, and prediction \cite{redmon2016yolo, redmon2018yolov3}. Collectively, these processes extract relevant features from input tensors, combine those features into a single representation, and then use that representation to predict the presence and location of objects within each tensor's unique coordinate system. By following these three steps, object detection neural networks can achieve cutting-edge performance for a variety of object detection applications.

The input tensors are analyzed in the feature extraction stage (or "the backbone") to identify relevant data patterns, typically using convolutional layers. A hierarchy of increasingly complex patterns is produced as the backbone processes the tensor \cite{krizhevsky2012imagenet, simonyan2015verydeep}. High-level estimator performance is still not assured, even though these extracted features are a more useful representation for predictions than raw data. In the feature aggregation stage, the object detection model will produce feature maps that are resistant to changes in scale, translation, and rotation \cite{tan2019efficientdet, lin2017feature, simonyan2015verydeep}. The feature aggregation stage is covered in detail in Section 2.B, Feature Aggregation. During the last stage, prediction (or “the neck”), aggregated features are transformed into an entire set of predictions. In computer vision’s object detection models, separate dense branches frequently perform classification and box coordinate regression for the detected objects \cite{redmon2016yolo, liu2016ssd}.

For computing final predictions in object detection, anchors, like those found in YOLO and SSD, have become standard \cite{redmon2016yolo, lin2017feature}. In order to localize objects in an image, anchors are a collection of predefined bounding boxes with various scales and aspect ratios. Anchors streamline the prediction process by splitting the task into two distinct tasks; determining whether an anchor box contains an object or not, and adjusting the anchor box coordinates to fit any present objects. This method enables the network to generalize objects of specific shapes and sizes while reducing the number of trainable parameters, speeding up parameter optimization. The network only predicts the presence and location of objects in a fixed set of boxes rather than predicting the precise location of each object across the entire image, making anchors computationally efficient. Anchors do not exist yet for SSL and sound detection models. However, feature aggregation is essential to proper function of anchors. Incorporating feature aggregation into sound detection architectures allows for the use of sound signal anchors in succeeding research, that is expected to substantially improve the localization and classification of sound sources at various amplitudes and phases \cite{liu2016ssd, redmon2016yolo, redmon2018yolov3}.

This study uses a control model consisting of only a basic backbone and head as a control model. The backbone of this control model consists of three sequential convolutions and the head has two branches, each with two sequential dense layers. As will be elaborated upon in the next section, Methodology, various feature aggregators are inserted between the backbone and neck of this control model; replicating the computer vision object detection architecture described in this section.

\section{Methodology} \label{sec:method}
\begin{figure}
    \centering
    \includegraphics[width=0.45\textwidth]{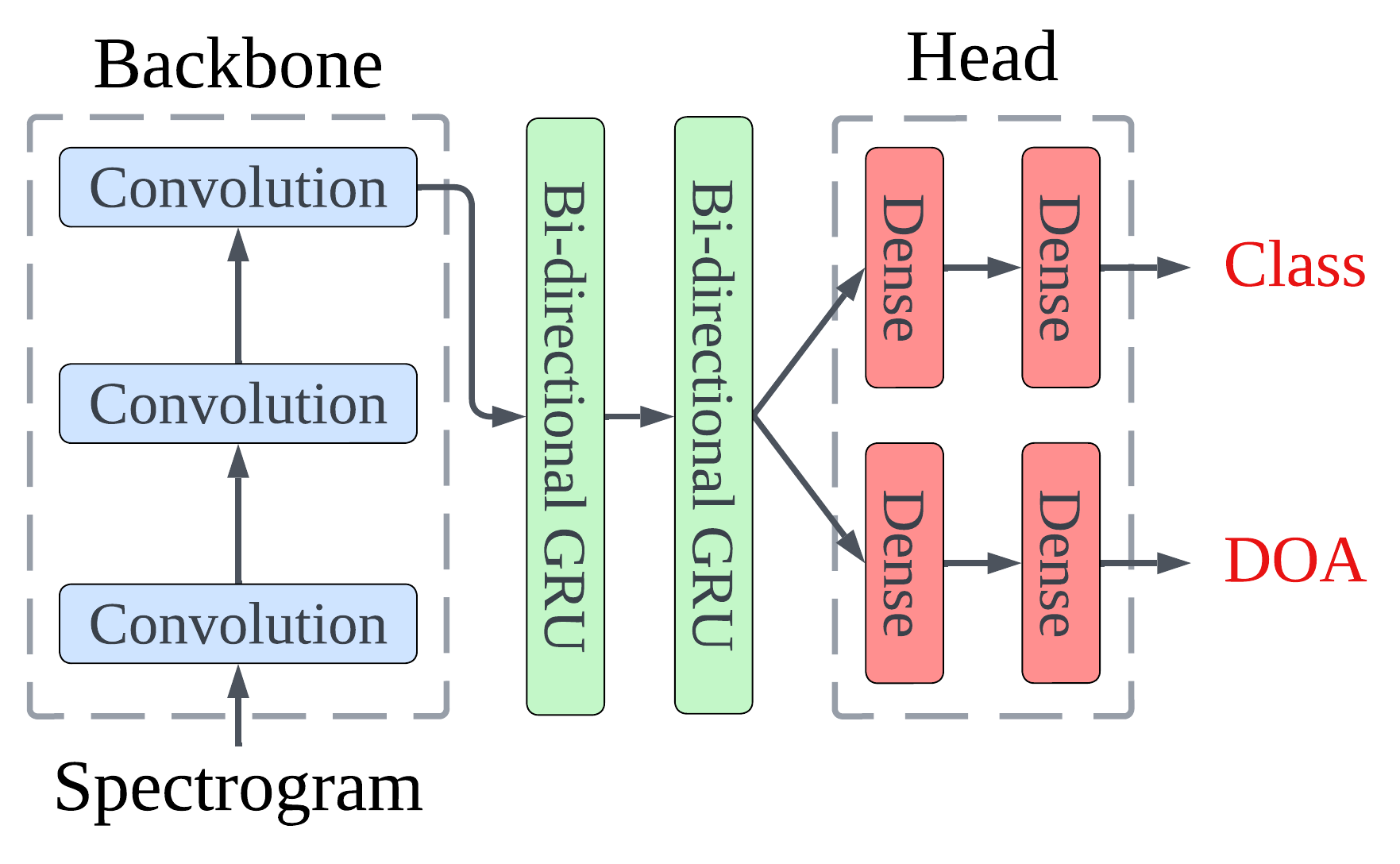}
    \caption{Illustration of the control model, SELDnet \cite{adavanne2018sound}.}
    \label{fig:SELD}
\end{figure}

To isolate the effects of feature aggregation on SSL models, this section will introduce four new SSL models that integrate various aggregators into a control model (taken from \cite{adavanne2018sound}) and then trained with identical datasets \cite{adavanne2018tut} and preprocessing \cite{adavanne2018sound}. The control model, SELDnet, can be seen in Fig. \ref{fig:SELD} and the proposed models with aggregation in Fig. \ref{fig:Full Models}.

\subsection{Development}
This study used Keras library with TensorFlow’s Functional API backend to implement and test the feature aggregation models. This API was chosen because of its flexibility for creating non-sequential neural networks. Feature aggregation nodes are Model class objects that incorporate sequential resampling, weighted averaging, and convolutional layers for processing input tensors. As the current version of the TensorFlow Functional API does not include a weighted averaging layer, we utilized a custom layer of the Layer class, with the weights designated as trainable variables. Subsequently, multiple node sub-models were interconnected to create Feature Aggregator sub-models, which were then integrated into the main model. This design allows nodes to be easily arranged to quickly create aggregators, and aggregators to be efficiently integrated into larger architectures.

Each model was trained for a max of 1000 epochs with early stoppage if the SELD score (see Section IV.C) on the test split does not improve for 100 epochs. This early stoppage is to prevent network over-fitting. For training loss, we utilized a weighted combination of binary cross-entropy for classification and MSE for localization with an Adam optimizer with default parameters \cite{kingma2015adam}.

Simple 1×1 convolutions were used in the feature aggregation nodes due to their ability to reduce dimensionality of feature maps \cite{szegedy2015goingdeeper} and apply nonlinear transformations \cite{goodfellow2016deep}. In terms of dimensionality reduction, the convolutional operation with a single filter and a stride of 1 in the time axis can be used to generate a new feature map with a reduced number of channels, which is particularly beneficial in deep neural networks where the number of feature maps can quickly become large and computationally expensive to process \cite{he2016deep}. As for nonlinear transformation, the convolutional operation with multiple filters and a nonlinear activation function, such as ReLu or softmax, are applied to feature maps to increase their expressive power and improve the performance of the neural network \cite{goodfellow2016deep}.

\subsection{Final Architectures}
The Feature Aggregators chosen for this study are PANet, BiFPN, and SEN. PANet and BiFPN were selected because of their well-established track record in popular object detection models, including YOLO and SSD \cite{tan2019efficientdet, liu2016ssd}. SEN is a new aggregator developed by this study. To evaluate the effects of the compression width value, two models with SEN are tested: one test model with a compression width of one (SEN\textsubscript{N=1}) and the other with a compression width of two (SEN\textsubscript{N=2}). The SEN model with a compression width of one uses two intermediate scales; these scales sizes are the averaged dimensions of their input tensors. All of these aggregators vary in number of nodes and connection patterns, allowing for analysis and speculation of optimal approaches for feature aggregation design.

The control architecture, SELDnet, is a MTL CRNN without feature aggregation. It simultaneously predicts the presence of multiple classes and their relative positions in 3D Cartesian coordinates. It has been chosen for a few reasons. First, this is currently a state-of-the-art architecture that performed well in a distinguished study by \cite{adavanne2018sound}. Second, the architecture design allows for easy integration of feature aggregators compared to non-sequential networks. Third, the control architecture’s hyperparameters have already been tuned, allowing this study to focus on tuning feature aggregators.

\begin{figure} [h]
   \centering
    \subfloat\centering (a){{\includegraphics[width=0.45\textwidth]{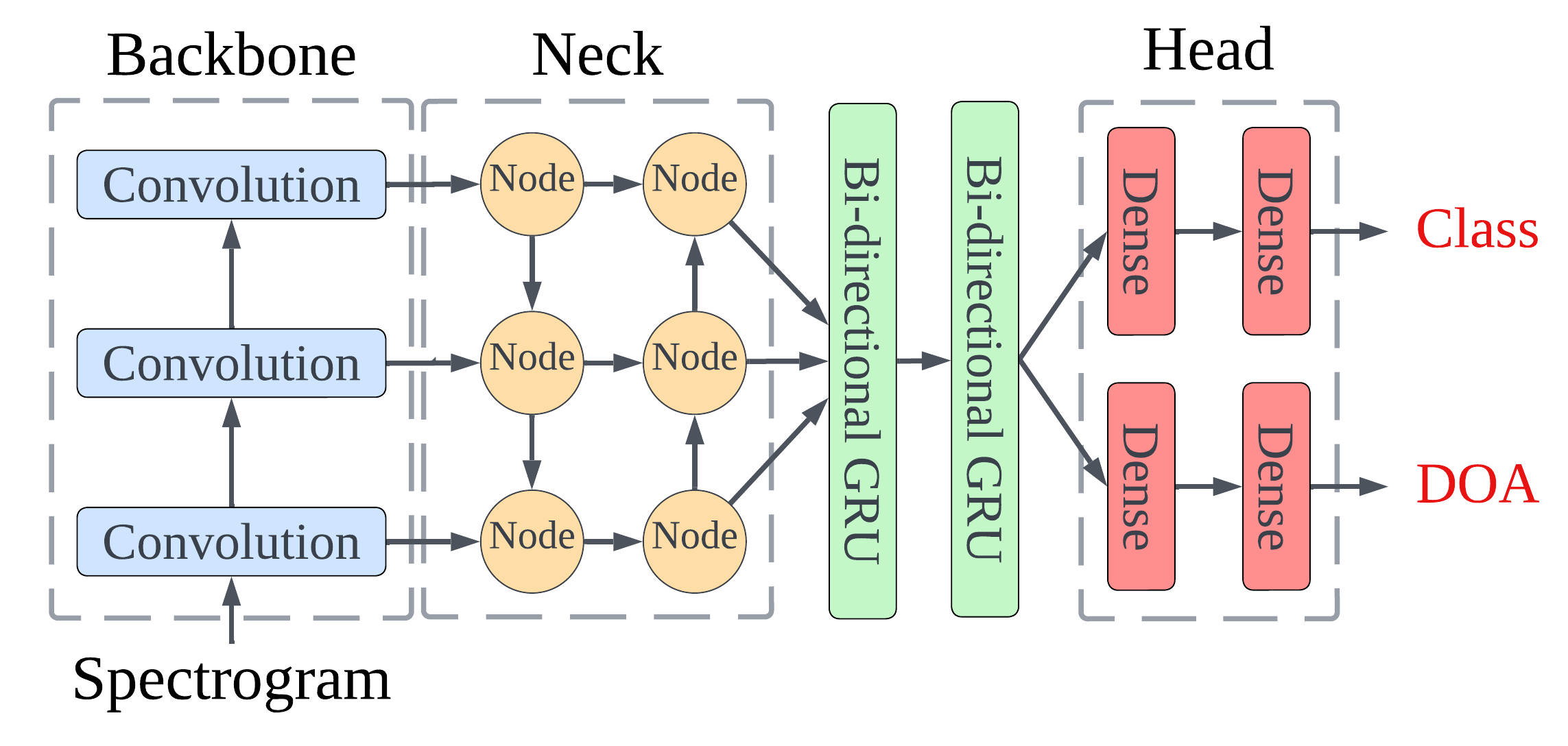}}}%
    \vspace{\baselineskip}
    
    \subfloat\centering (b){{\includegraphics[width=0.45\textwidth]{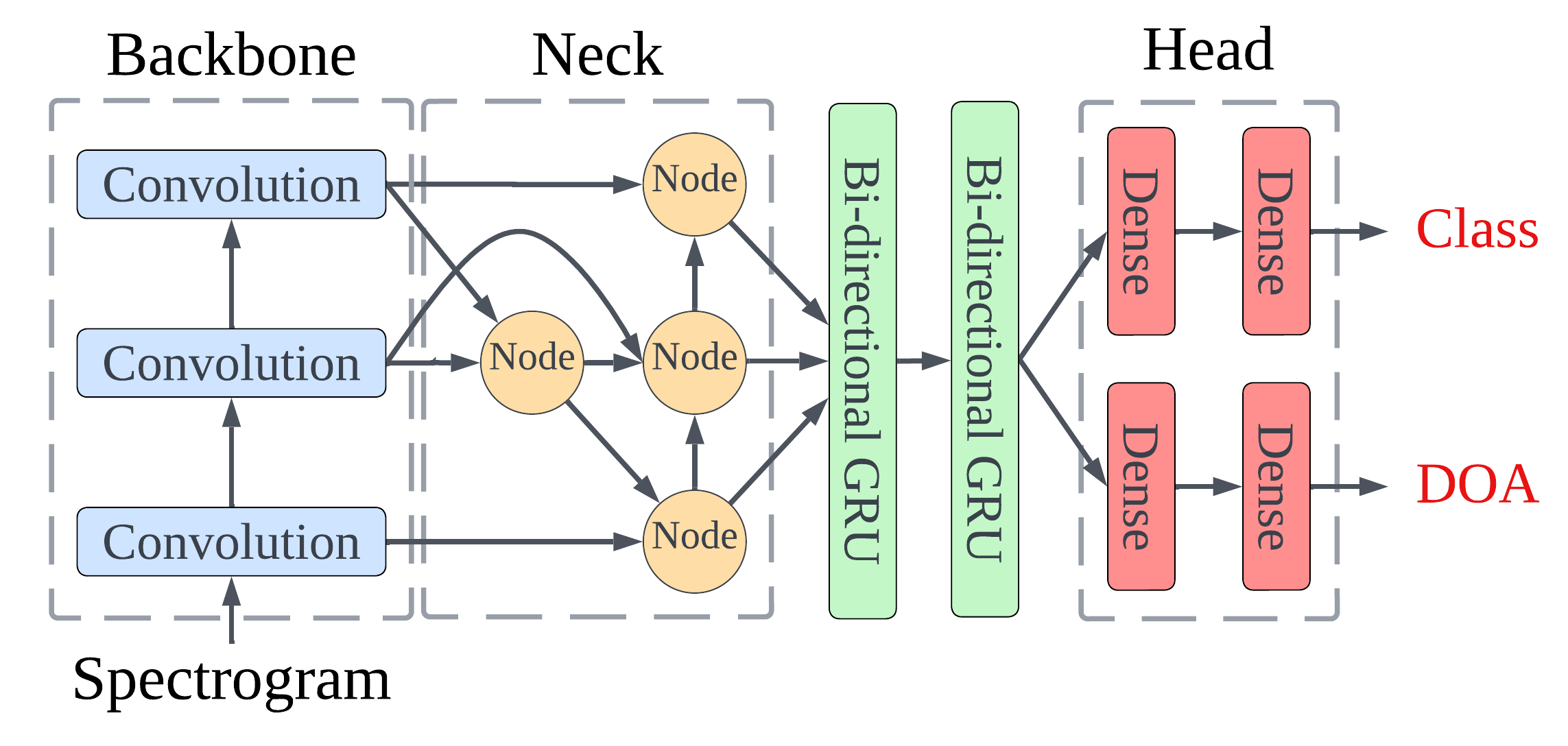}}}%
    \vspace{\baselineskip}
    
    \subfloat\centering (c)
    {{\includegraphics[width=0.45\textwidth]{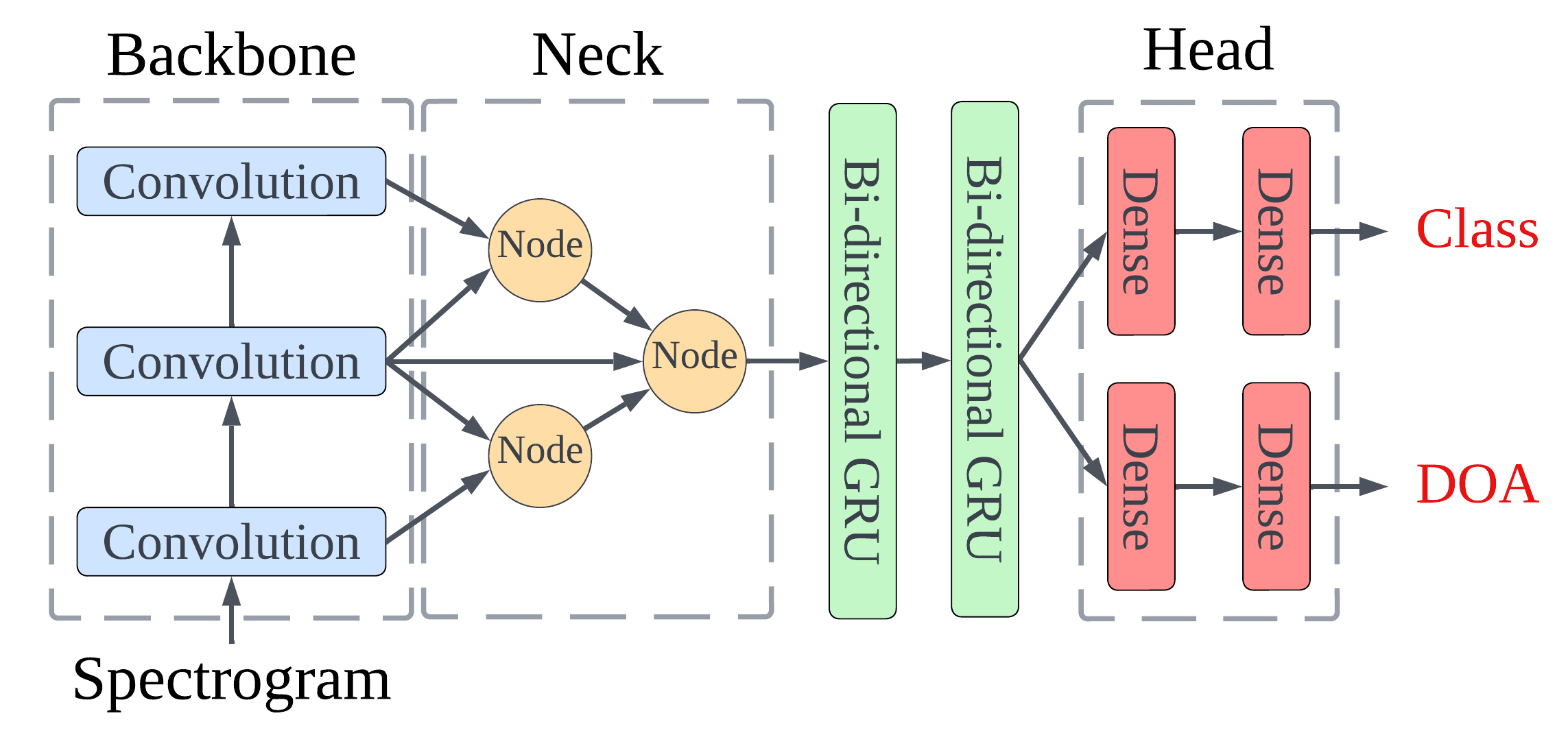}}}%
    \vspace{\baselineskip}
    
    \subfloat\centering (d)
    {{\includegraphics[width=0.45\textwidth]{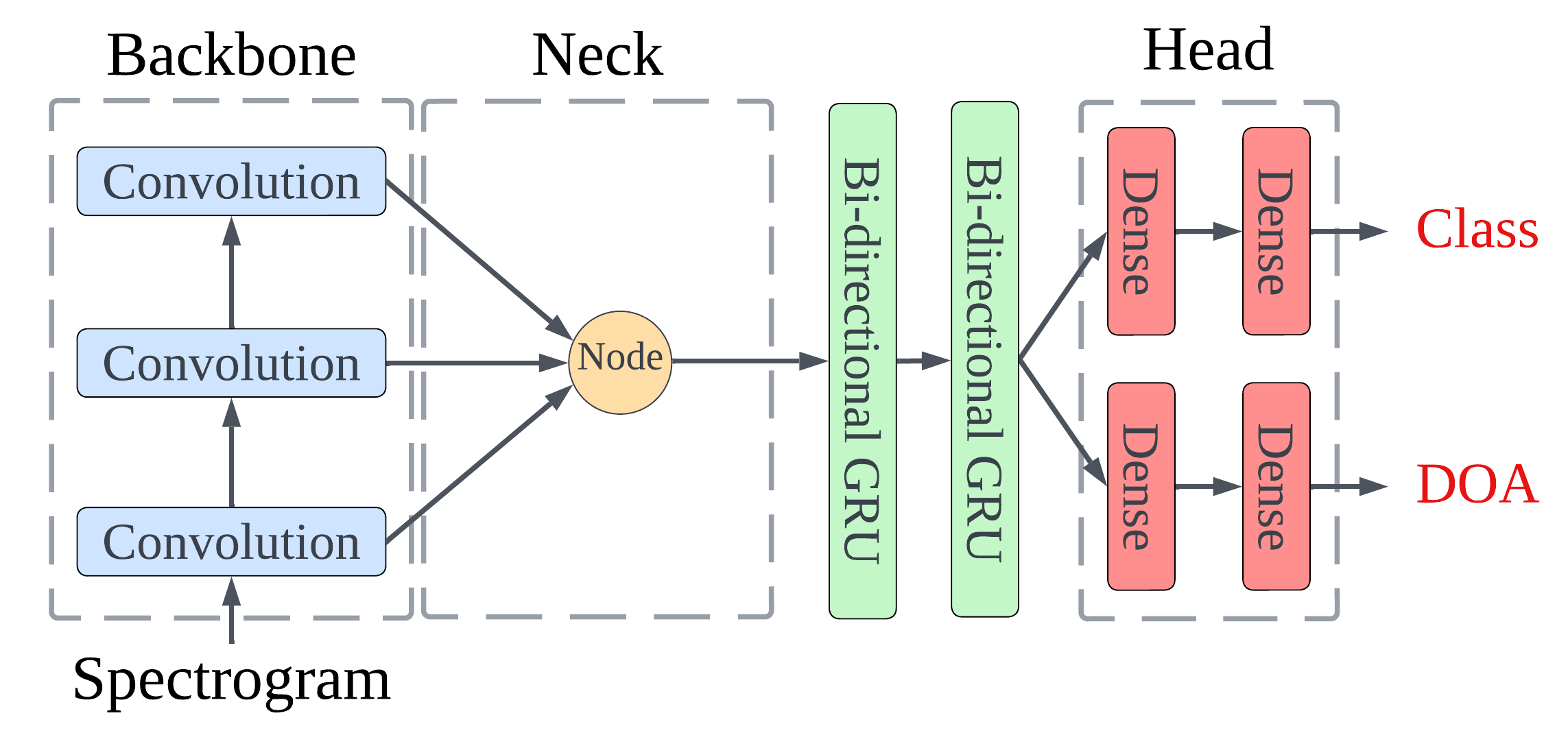}}}%
    \vspace{\baselineskip}
    
    \caption{Diagrams of final model architectures proposed by this study. Subfigures a, b, c, and d illustrate SELDnet with PANet, BiFPN, SEN\textsubscript{N=1}, and SEN\textsubscript{N=2}, respectively.}
    \label{fig:Full Models}
\end{figure}

As observed in Fig. \ref{fig:SELD}, the control model’s feature extraction and prediction stages are completed by three sequential convolutions and two branches of two dense layers, respectively. Fig. \ref{fig:Full Models} presents the proposed architectures by this study. Figs. \ref{fig:Full Models} (a)-(b) display SELDnet with the established PANet and BiFPN aggregators. Figs. \ref{fig:Full Models} (c)-(d) illustrate SELDnet with two variations of SEN. The first variation, Fig. \ref{fig:Full Models} (c), incorporates an aggregator with two SEN layers with compression width of one. The second variation, Fig. \ref{fig:Full Models} (d), demonstrates the SELDnet with a single SEN layer of compression width two.

\section{Evaluation} \label{sec:eval}
\subsection{Dataset}
The REAL dataset, compiled by \cite{adavanne2018tut}, serves as a valuable resource for research on sound event detection (SED) and localization \cite{adavanne2018sound}. The dataset consists of 216 uncompressed WAV audio recordings, each lasting 30 seconds, captured in various indoor and outdoor settings such as street junctions, tram stations, shopping malls, and pedestrian streets. These settings represent common acoustic environments found in urban areas and feature diverse overlapping sound sources and backgrounds.

The dataset includes spatial coordinates for the loudspeakers and microphones, along with annotations that provide information about the temporal boundaries, classification, and spatial coordinates of sound events present in each recording. It encompasses 11 distinct sound categories, including car, bus, train, tram, footsteps, speech, music, dog, bird, jackhammer, and siren.

\subsection{Preprocessing}
In order to discern the specific impact of feature aggregation within our framework, the data preprocessing methodology employed is identical to the control model study \cite{adavanne2018sound} and uses the code from this reference repository. The preprocessing stage involves the following steps:
\vspace{\baselineskip}

\begin{enumerate}[itemsep=10pt]
  \item Raw audio files were de-noised using band-pass filtration to remove low and high-frequency noise. This method is effective against typical noise in recording environments \cite{adavanne2018sound}. Signals were then downsampled to 16 kHz, reducing computational complexity and ensuring efficient data preprocessing for future stages.
  
  \item A Short-Time Fourier Transform (STFT) extracted features from the preprocessed audio signals to create a detailed time-frequency representation \cite{xie2019doa}.  A window size of 1024 samples and hop size of 256 samples provided the optimal temporal and spectral details in the resulting spectrogram \cite{yin2018doa}. This spectrogram was used as the neural network input for SED and localization.
  
  \item To bolster the training dataset and improve model generalization, data augmentation techniques were employed, including random time shifting, frequency shifting, and amplitude scaling \cite{han2020feature}. This enriched dataset enhanced the model's adaptability and performance in varied acoustic scenarios.
\end{enumerate}

\subsection{Metrics}
Each model's performance is evaluated using several metrics that measure the accuracy of the SED and the sound event DOA estimation. The SED metrics are F-score and Error Rate. F-score ($F$) is a widely used metric for binary classification problems that measures the balance between precision and recall \cite{powers2011evaluation}. It is defined as the harmonic mean of precision and recall. In the context of SED, True Positives ($TP$) refer to the correctly detected events, False Positives ($FP$) refer to the events that were incorrectly detected, and False Negatives ($FN$) refer to the events that were missed by the model \cite{adavanne2018sound}. F-score is a useful metric because it considers both the number of correctly detected events and the number of missed and false alarms. A higher F-score indicates a better performance. The F-score for predicting the presence of classes in each k one-second segment is defined as 
\begin{equation}
F = \frac{2 \cdot \sum_{k=1}^{K} TP(k)}{2 \cdot \sum_{k=1}^{K} TP(k) + \sum_{k=1}^{K} FP(k) + \sum_{k=1}^{K} FN(k)}
\end{equation}
where, for each one-second segment $k$, $TP(k)$ is the number of true positives; $FP(k)$ is the number of false positives; and $FN(k)$ is the number of false negatives. True positives are correctly predicted sound event classes which are present in the segment. In contrast, false positives are sound event classes erroneously predicted to be within the segment and false negatives are sound event classes present in the segment but failed to be detected.

Error Rate (ER) is another commonly used metric for SED, which measures the percentage of incorrectly detected events \cite{adavanne2018sound, mesaros2016metrics}. ER is calculated as:
\begin{equation}
ER = \frac{\sum_{k=1}^{K} S(k) + \sum_{k=1}^{K} D(k) + \sum_{k=1}^{K} I(k)}{\sum_{k=1}^{K} N(k)}
\end{equation}
where, for each one-second segment $k$, $N(k)$ is the total number of active sound event classes in the ground truth. $S(k)$, substitution, is the number of times an event was detected at the wrong level and is calculated by merging false negatives and false positives without individually correlating which false positive substitutes which false negative. The remaining false positives and false negatives, if any, are counted as insertions $I(k)$ and deletions $D(k)$, respectively. These values are calculated as follows:
\begin{align}
S(k) {}={}& min(FN(k), FP(k))\\
D(k) {}={}& max(0, FN(k)-FP(k))\\
I(k) {}={}& max(0, FP(k)- FN(k))
\end{align}

The DOA metrics are DOA error and Frame Recall. DOA error measures the difference between the estimated DOA and the ground truth DOA in degrees for the entire dataset with total number of DOA estimates, $D$ \cite{adavanne2018sound}. A lower DOA error indicates a better performance. The error is defined as
\begin{equation}
DOA\hspace{0.05in}Error = \frac{1}{D} \sum_{d=1}^{D} \sigma((x_G^d,y_G^d,z_G^d),(x_E^d,y_E^d,z_E^d))
\end{equation}
where $(x_E,y_E,z_E)$ is the predicted DOA estimate, $(x_G,y_G,z_G)$ is the ground truth DOA, and $\sigma$ is the angle between $(x_E,y_E,z_E)$ and $(x_G,y_G,z_G)$ at the origin for the $d$-th estimate:
\begin{equation}
\sigma = 2 \cdot \arcsin\left(\frac{\sqrt{\Delta x^2 + \Delta y^2 + \Delta z^2}}{2}\right) \cdot \frac{180}{\pi}
\end{equation}
with $\Delta x=x_G-x_E$,  $\Delta y=y_G-y_E$, and $\Delta z=z_G-z_E$.

Frame recall, as defined by \cite{adavanne2018sound}, is a metric used to measure the accuracy of a model's predictions in the context of time frames or segments. It measures the percentage of time frames where the number of estimated and ground truth DOAs are unequal. A higher Frame Recall indicates a better performance and is calculated as:
\begin{equation}
FR = \frac{\sum_{k=1}^{K} TP(k)}{\sum_{k=1}^{K} TP(k) + FN(k)} \cdot 100
\end{equation}

We used a combined localization and classification score, SELD, to perform early training stoppage. If the SELD score did not improve over 100 epochs, training was terminated to prevent overfitting. SED and DOA scores represent the overall performance of an estimator for sound event detection and localization, respectively. SELD is the average of these scores and functions as a single overarching metric to compare models. A lower value indicates better performance for DOA, SED, and SELD scores. Eqs. \eqref{DOA Score}, \eqref{SED Score} and \eqref{SELD} define these metrics \cite{adavanne2018sound}:
\begin{align}
\label{DOA Score}
\text{\textit{DOA score}} {}={}& \frac{\left(\text{\textit{DOA Error}}/180 + \left(1 - FR/100\right)\right)}{2}\\
\label{SED Score}
\text{\textit{SED score}} {}={}& \frac{\left(ER + \left(1 - F/100\right)\right)}{2}\\
\label{SELD}
\text{\textit{SELD}} {}={}& \frac{\left(\text{\textit{SED score}} + \text{\textit{DOA score}}\right)}{2}
\end{align}

\subsection{Baseline Methods}
SELDnet is an advanced deep learning architecture developed specifically for combined SED and DOA estimation tasks \cite{adavanne2018sound}. SELDnet processes spectral features dedicated to the SED task in parallel with spatial features for the DOA estimation. By leveraging the fusion of convolutional and recurrent layers, SELDnet effectively pinpoints both the occurrence and the spatial origin of a sound event. MSEDnet, derived from SELDnet, is designed for monaural (single-channel) SED. By focusing on the SED task, MSEDnet provides an optimal solution for applications where the sole requirement is event detection without the need for spatial localization \cite{adavanne2017sound}. Its architecture is tailored to single-channel audio contexts, ensuring precise event detection. SEDnet stands as a dedicated solution for SED, capitalizing on deep learning techniques \cite{adavanne2017sound}. With its architecture centered around event identification, SEDnet excels in scenarios where temporal detection of sound events is the primary objective. It delivers accuracy and efficiency in sound event classification without incorporating spatial estimation components. DOAnet offers a specialized approach towards the spatial dimension of audio signals, focusing solely on DOA estimation \cite{adavanne2018direction}. MUSIC (Multiple Signal Classification) is a robust algorithm for DOA estimation. Relying on subspace methods, MUSIC differentiates the signal space from the noise space, facilitating precise DOA predictions for multiple sound sources \cite{schmidt1986multiple}. Its mathematical foundation and proven efficiency in array signal processing render it as a reliable choice for DOA estimation tasks, even in contexts dominated by neural network models \cite{adavanne2018sound}.

\section{Results}\label{sec:res}
The results in TABLE \ref{table:Results Table} indicate that feature aggregation enhanced the control model's capacity in both localization and classification of the sound sources. None of the models evaluated in this study outperformed the algorithms specialized for only classification or localization, but all demonstrated a clear improvement in both tasks compared to the control model, SELDnet. 

\captionsetup{justification=centering}

\begin{table*}
\centering
\caption{\\\hspace{\textwidth}CLASSIFICATION AND LOCALIZATION SCORES OF TEST ARCHITECTURES COMPARED TO CONTROL ALGORITHMS.}
\label{table:Results_Table}
\scalebox{1.3}{
    \setlength{\extrarowheight}{10pt}
    \begin{tabular}{|c|c|c|c|c|c|c|c|}
    \hline
    \multirow{2}{*}{Architecture \textbackslash\ Algorithm} & \multicolumn{3}{c}{Classification} & \multicolumn{3}{|c|}{Localization} & Joint Task \\
    \cline{2-8}
    & ER & F-Score & SED & DOA Error & Frame Recall & DOA Score & SELD Score\\
    \hline
    SELDnet \cite{adavanne2018sound} & 0.41 & 60.5 & 0.40 & 26.9 & 65.3 & 0.25 & 0.325 \\
    SELDnet + PANet & 0.36 & 65.1 & 0.35 & 11.8 & 72.5 & 0.17 & 0.262 \\
    SELDnet + BiFPN & 0.37 & 63.0 & 0.37 & 15.2 & 68.6 & 0.20 & 0.285 \\
    SELDnet + SEN\textsubscript{N=1} & 0.36 & 63.6 & 0.36 & 14.3 & 69.5 & 0.19 & 0.277 \\
    SELDnet + SEN\textsubscript{N=2} & 0.37 & 62.5 & 0.37 & 16.1 & 68.0 & 0.20 & 0.289 \\
    MSEDnet \cite{adavanne2017sound} & 0.35 & 66.2 & 0.34 & - & - & - & - \\
    SEDnet \cite{adavanne2017sound} & 0.38 & 64.6 & 0.37 & - & - & - & - \\
    DOAnet \cite{adavanne2018direction} & - & - & - & 6.30 & 46.5 & 0.29 & - \\
    MUSIC \cite{schmidt1986multiple} & - & - & - & 36.3 & - & - & - \\
    \hline
    \end{tabular}
}
\end{table*}

\subsection{Classification}
For the SELDnet variants, classification improvements were clear but minimal. All of the ER scores are closely clustered, making it difficult to determine the extent to which different feature aggregation designs affected scores. However, the F and SED scores imply that aggregation did improve SELDnet’s ability to classify in a manner comparable to MSEDnet and SEDnet. The dataset may impose restrictions on these scores, preventing architectural design from displaying a large impact. Datasets can limit deep learning model performance through factors such as data size, quality, class imbalances, noisy or biased labeling, and distribution mismatches; all of which hinder the model's ability to achieve over a certain score. Compared to MSEDnet, both SELDnet+PANet and SELDnet+SEN\textsubscript{N=1} performed marginally worse. These two models performed marginally better than SEDnet, whereas SELDnet with BiFPN and SEN\textsubscript{N=2} performed comparably to SEDnet. With respect to SEDnet, SELDnet with BiFPN and SEN\textsubscript{N=2} scored slightly better on ER, slightly worse on F-Score, but achieved the same overall SED score. Overall compared to the control model, the models with aggregation provide better performance; demonstrated by lower ER, higher F-Score, and higher SED score.

\subsection{Localization}
The improvements in localization scores of models with aggregation are notable. Compared to SELDnet, all the networks with aggregation considerably improved performance, as was expected. All aggregated models had considerably lower DOA Errors, higher Frame Recall, and lower DOA score. SELD+PANet performed particularly well in DOA estimation, with metric scores that stand out from the other clustered aggregation DOA scores. This clear DOA estimation boost suggests that feature aggregation enhances the distinguishing of various sound signals, such as reverberations, diffractions, and direct signals. The increased robustness to indirect signals, observed after introducing aggregators into SELDnet, is attributable to the enhanced feature scaling. Indirect signals, such as reflections, can exhibit comparable wave patterns to direct signals at lesser amplitudes. Therefore, one would anticipate that a better comprehension of feature scales would enhance the differentiation between direct and indirect signals.

Compared to DOAnet, SELDnet and its variants have a higher frame recall and overall DOA Error; signifying that they excel at correctly identifying DOAs within individual time frames while maintaining consistency with the ground truth DOAs. This indicates that these models have difficulty minimizing the overall disparity between predicted and ground truth DOA compared to DOAnet, but consistently capture signal patterns with respect to time.

It is likely that DOAnet’s inconsistency with respect to time is a result of an inability to distinguish direct signals from reflections, reverberations, and diffractions. The data implies that SELDnet variants (developed by this study) are adept at pinpointing the precise time instances when sound sources appear, leading to an improved ability to distinguish between multiple signals. Furthermore, by consistently capturing signal patterns over time, these models are likely to be more robust in dynamic soundscapes where the number of sound sources and noise interference can vary. This trait is vital in real-world applications where sound sources often overlap and vary in number and characteristics.

\subsection{Joint Classification and Localization}
The SELD scores indicate that, regardless of aggregator design, feature aggregation improves the function of joint sound classification and localization models. Although aggregators with more nodes outperformed the single node SEN model, this modest aggregator demonstrated that even minimal aggregation can counter-act the negative effects of the semantic gap. Clearly, PANet’s in-depth and equal processing of all scales is optimal for performance. However, as previously discussed, this approach can be computationally demanding, which may prove adverse for certain situations.

\subsection{Aggregator Comparison}
This section will compare aggregators using two metrics: their overall percentage improvement on SED, DOA, and SELD scores and the percentage improvement per node. The percentage improvement per node is intended as a metric to quantify the efficiency of aggregator designs. As can be seen in TABLE \ref{table:Aggregator Comparison}, although some aggregators have better overall improvement, others have a better improvement ratio per node, implying a more efficient connection design.

\begin{table*}
  \centering
  \caption{\\\hspace{\textwidth}COMPARISON OF AGGREGATORS’ EFFECTS ON CONTROL MODEL SCORES.}
  \scalebox{1.3}{
    \setlength{\extrarowheight}{6pt}
      \begin{tabular}{|c|c|c|c|c|c|c|c|}
        \hline
        \multirow{2}{*}{Aggregator} & \multirow{2}{*}{\# Nodes} & \multicolumn{2}{c|}{SED \% Improvement} & \multicolumn{2}{c|}{DOA \% Improvement} & \multicolumn{2}{c|}{SELD \% Improvement} \\
        \cline{3-8}
        & & Overall & Per Node & Overall & Per Node & Overall & Per Node \\
        \hline
        PANet & 6 & 12.5 & 2.08 & 32.0 & 5.33 & 19.4 & 3.23 \\
        BiFPN & 4 & 7.50 & 1.88 & 20.0 & 5.00 & 12.3 & 3.08 \\
        SEN\textsubscript{N=1} & 3 & 10 & 3.33 & 24.0 & 8.00 & 14.8 & 4.92 \\
        SEN\textsubscript{N=2} & 1 & 7.50 & 7.5 & 20.0 & 20.0 & 11.1 & 11.1 \\
        \hline
        \end{tabular}
    }
  \label{table:Aggregator Comparison}
\end{table*}

The obvious outlier is SEN\textsubscript{N=2}. This model’s results imply that any aggregation helps counteract the semantic gap. The collective results indicate that less nodes result in a higher percentage improvement per node. However, in this particular SEN\textsubscript{N=2} aggregator, the magnitude of improvement per node must be taken with a grain of salt due to the simplicity of the aggregator. SELDnet is an unusually compact neural network, and most real-world models are much deeper (such as Darknet 53 with a backbone of 53 convolutional layers). For  backbones deeper than three layers, which is the case for most models, SEN aggregators with a compression width of two would involve more than one node. We hypothesize that the use of a single node causes this SEN\textsubscript{N=2} to seem disproportionally effective per node because the overall score change is divided by one. The overall improvement from this SEN\textsubscript{N=2} aggregator is a testament to the effects of having any aggregator, but the results for improvement per node are skewed due to division by one. Nevertheless it is interesting that compared to
BiFPN, this single node performed comparatively in overall SED and DOA percentage improvement and slightly worse in overall SELD. It is important to note that the DOA and SED scores of BiFPN and SEN\textsubscript{N=2} are the same because of rounding, but the actual difference is seen in the SELD score. As will be discussed later in this section, we attribute this similar performance to the efficacy of encoder aggregators of SEN.

After removing this outlier and comparing PANet, BiFPN and SEN\textsubscript{N=1}, the next clear take away is PANet’s overall percentage improvement. PANet was not as efficient as SEN\textsubscript{N=1}, but the overall improvement is substantial compared to all other aggregators. This is attributable to the in-depth processing at every scale, which is the most comprehensive design for addressing the semantic gap. The lower efficiency per node is likely the result of certain scales not requiring as much processing as is actively occurring.

SEN\textsubscript{N=1} is clearly an efficient design, with the second highest overall percentage improvements and highest percentage improvement per node, when excluding the outlier SEN\textsubscript{N=2}. This efficiency per node is attributable to the efficacy of the encoder approach. This approach allows for even weighing and consideration of all scales, like PANet, without leading to uneven processing of each scale, as seen in BiFPN. This equal evaluation of all scales with reduced nodes leads to an efficient aggregation process. BiFPN’s results indicate that the design is not the best performing or most efficient. Architectures with more trainable parameters, such as the additional nodes in PANet, can be trained to outperform BiFPN on the test dataset. It is interesting that the SEN models performed well in comparison with BiFPN, which has more nodes than both SEN models. We hypothesize that BiFPN overemphasized one scale due to the scale's extra nodes, whereas SEN created a compressed representation with equal weighting of all scales.

\section{Conclusion}\label{sec:conc}
The results indicate that, regardless of the aggregator's design, feature aggregation can significantly improve the performance of neural networks for sound detection. A balance must be struck between computational expense and performance when deciding on an aggregator. While examining the available aggregators, SEN and PANet stand out as the most cost-effective and robust, respectively. The difference in aggregator performances indicates that when performing feature aggregation, it is best to equally emphasize all scales. Future research may delve deeper into an assortment of topics, such as the establishment of anchors for spectrograms and the development of more complex SEN designs (such as stacking SEN after FPN or PANet).

\section*{Acknowledgment}
We would like to express our gratitude to Tampere University of Technology and its licensors for granting permission to use the code for the Sound Event Localization and Detection using Convolutional Recurrent Neural Network method/architecture, which is available in the GitHub repository with the handle "seld-net" found at https://github.com/sharathadavanne/seld-net. This code was described in the paper titled "Sound event localization and detection of overlapping sources using convolutional recurrent neural network" \cite{adavanne2018sound}. 

We acknowledge and honor the non-commercial nature of this grant and affirm our commitment to preserving the copyright notice in all reproductions of this work. Furthermore, we are grateful to the original source of the work, the Audio Research Group, Lab. of Signal Processing at Tampere University of Technology.

\ifCLASSOPTIONcaptionsoff
  \newpage
\fi

\bibliographystyle{IEEEtran}
\bibliography{references}

\end{document}